# Kinetic processes in inhomogeneous self-aware media

A. Morozovskiy[1], A.A. Snarskii[2], I.V. Bezsudnov[3], V.A. Sevryukov[3], J. Malinsky[4]

[1] Corresponding author. E-mail: alexmor@gmail.com
[2] Dep. of General and Theoretical Physics, National technical university of Ukraine "KPI"
[3] ZAO "NPP Nauka-Service", Moscow, Russia
[4] CUNY Graduate Center, Physics Program, BCC, Department of Physics and Technology

## Abstract

The new framework for finance is proposed. This framework based on three known approaches in econophysics. Assumptions of the framework are the following:

- For the majority of situations market follows non-arbitrage condition.
- For the small number of situations market influenced by the actions of big firms.
- If actions of big players lead to the arbitrage opportunity, small players could self-organize to take advantage of this opportunity.

Suggested framework is applied for the analysis of market impact models, behavior of big players, self-organization of market firms and volatility description.

Category: Statistical Mechanics (cond-mat.stat-mech)

Comment: The presented material is the Chapter 23 of the book "Transport processes in macroscopically disordered media (From mean field theory to percolation)" to be published by Springer ISBN 978-1-4419-8290-2.

32 pages, 1 figure, 2 tables

# Contents

# Introduction

The kinetic processes described so far were connected with phenomena, existing in nature. There is also another kind of kinetic processes – the ones happening in human society. In what follows we will consider only one example of such processes – processes describing changes of values of financial instruments (for example, changes in prices of stocks). The main differences between these 2 groups of kinetic processes (natural and social) is that participants in the latter group are self-aware – if they could deduce the result of the kinetic process, they could change this result in a such way that result would be beneficial for them. The existence of this difference, however, is not necessary a reason for abandoning the possibility of application of methods of natural sciences to social phenomena, it is only necessary to keep in mind an existence of a difference.

Recently, a new discipline appeared – econophysics (Mantegna and Stanley 1999). This discipline describes behaviour of financial markets using physical methods. It could be said, however, that economists also used physical methods (ordinary and stochastic differential equations). For us the main difference between economics and econophysics is the difference between people, who work in the corresponding areas (economists and physicists). It could be argued that in contrast to economists, who are using mostly methods of physics of the nineteen and the beginning of the twentieth century, physicists are using also physical methods developed in the middle and at the end of the twentieth century (advances in statistical physics, critical phenomena, fractals).

Particularly, there are big breakthroughs that had happened in econophysics. They will be described not in order of their appearance, but in the order of their strength (from smaller to larger market moves):

1. Work of Bouchaud and al. (Bouchaud et al. 2009), describing market impact models,

2. Work of Stanley and al. (Gabaix et al. 2003), describing behavior of big players in the market,

3. Work of Sornette and al. (Sornette and Johansen 2001), describing a self-organizing behaviour of the market during market crashes.

It could be said, that as in a Greek mythology, these models described behaviour of different participants of the market: simple people, kings and gods: if the work of Bouchaud and al. describes a behaviour of every day functionality of market ("simple people"), and work of Stanley and al. describes a behaviour of big players ("kings"), the work of Sornette and al. describes the behavour of market during crashes ("gods").

# 1. Stylized facts

Economics is an enormous area of life. There are many different phenomena and facts related to the economics. Because the area of experience in economics is enormous, it is reasonable to create a simple set (dictionary) of well represented facts. Such collection of facts called stylized facts. Even the literature on stylized facts is enormous (see, for example (Lux 2007)). Only small number of classified stylized facts will be discussed here.

The following stylized facts are will be basis for the following discussion:
- Absence of correlation between returns (in average).
- Power laws for different quantities.
- Power dependence of order flow.
- Volatility clustering.
- Existence of crashes and formulas describing them.

All of these facts will be discussed below in more details.

**Absence of correlation between returns (in average).**
If there is a correlation in returns, this correlation could be exploited by discoverers of the correlation (for profit) and eventually this correlation will disappear. This stylized fact is considered being a principle, and it is a most important principle in the modern finance.

**Power laws for different quantities.**
Following (Gabaix et al. 2003), let's define different quantities and describe their empirical distribution. If we define return $rt$ as a change in logarithm of price for a time interval $\Delta t$ ($p(t)$ – price of stock at time $t$): $rt = lnp(t) - lnp(t - \Delta t)$, than it was found (Gabaix et al. 2003) that probability of return for the value $|r(t)| > x$ could be described as:

$$P(|r_t| > x) \sim x^{-\xi_r} \quad , \text{ where } \xi_r \approx 3 \tag{1.1}$$

According to (Gabaix et al. 2003), similar laws with different indices exist for 3 additional quantities: distribution of trading volume, number of trades, and the market values of managed assets.

For trading volume $V$:

$$P(V_t > x) \sim x^{-\xi_V}, \quad \text{where } \xi_V \approx 1.5 \tag{1.2}$$

For number of trades $N_t$:

$$P(N_t > x) \sim x^{-\xi_N}, \quad \text{where } \xi_N \approx 3.4 \tag{1.3}$$

For market value of managed assets $S$:

$$P(S > x) \sim x^{-\xi_S}, \quad \text{where } \xi_S \approx 1.05 \tag{1.4}$$

**Power dependence of order flow.**

It was observed (Bouchaud et al. 2009) that there is a power law correlation between signs of the trade (if buy is plus, and sell is minus). It could be defined as long-memory of order flow.

**Volatility clustering.**

It is observed, that volatility is correlated – specifically big returns follow by big returns, and small returns by small ones.

**Existence of crashes and formulas describing them.**

Well developed theory of crashes exists in the financial economics. The models (Sornette and Johansen 2001) have been developed in order to quantify stock dependence on time.

# 2. Description of existing models

## *2.1. Model of the market impact*

One of extraordinary results obtained during development of econophysics was obtained by Bouchaud, Farmer, Lillo and al. (Bouchaud et al. 2009). In this review and in articles mentioned in it, very simple and elegant condition for market impact was derived. Let's repeat their arguments. Specifically, let's assume that there is an expected direction of the price change for the stock $S$. Then, if probability of price change is $p$ and expected price change is $\Delta S_1$, then the following formula would hold:

$$p\Delta S_1 + (1-p)\Delta S_2 = 0, \tag{2.1}$$

where $\Delta S_2$ – is unexpected price change. This is a non-arbitrage equation for one tick and it makes sense. Particularly, when $p$ is close to 1 and everybody expects small change in one direction, there is a very small probability of big unexpected move $\Delta S_2$ in the opposite direction.

As it was discussed earlier this stylized fact is connected with order flow: order flow is correlated in time. It was shown in (Bouchaud et al. 2009) that fixed permanent impact model is not compatible with these stylized fact. In order to create model that incorporates order flow dependence over time and market efficiency 2 different approaches were proposed (see (Bouchaud et al. 2009)): transient impact framework, and history dependent, permanent impact. It is also shown in (Bouchaud et al. 2009) that these approaches are equivalent.

## *2.2. Description of the dynamics of big impact players*

Now, let's describe dynamics of big players, as it provided in (Gabaix et al. 2003). It was observed that there are critical indexes for distributions of different quantities (returns, volumes, sizes).

The values of these indexes are provided in the section "Stylized facts". It appears that there is a relation between return and volume for big players. We will start with derivation of such formula.

### *2.3. Derivation of relation between return and volume*

It is assumed in the framework suggested by (Gabaix et al. 2003) that dependence between return and volume is determined by the behaviour of big players in the market, for example, fund managers. The idea of derivation is based on the optimality of the strategy for the big players (they lose the minimal possible amount of money). The strategy, basically, should provide minimal front-running for the observers of markets.

Let's consider the situation of the fund manager. It is assumed here that the fund manager decided that the value of the stock (or the other instrument) is divergent from "true price". If he or she decides to buy or sell this instrument, he or she should be sure that his action would not move the market. Mathematically it could be expressed in the following way below. Buying $V$ shares would lead to the price increase $\Delta p$ and relative price increase $\frac{\Delta p}{p_0}$ (where $p_0$ – initial price of instrument). If individual seller provide $s$ shares, the total number of shares provided by all sellers (by all market), could be written as :

$$q = s\frac{\Delta p}{p_0} \qquad (2.2)$$

The more fund manager waited for available shares, the more of them become available. So, in time $T$ manager could have the following number of shares $N_1$:

$$N_1 = kTq \qquad \text{or} \qquad N_1 = kTs\frac{\Delta p}{p_0} \qquad (2.3)$$

The process of acquiring shares would stop when $N_1$ would become equal to $V$:

$$N_1 = V \qquad (2.4)$$

The time to this moment is proportional to:

$$T \sim \frac{V}{\Delta p} \qquad (2.5)$$

The equation (2.5) is logical. The bigger price impact of trading, the less time is needed to wait for buying $V$ shares, and backwards, the less time manager spent on buying shares, the bigger price impact:

$$\Delta p \sim \frac{V}{T} \qquad (2.6)$$

In order to formulate manager problem 2 other quantities are also important. One of these quantities is mispricing: $M$ (it is assumed here that manager of the fund correctly identified mispricing of the share). Another of the quantity is $\mu$: the speed with which mispricing is eliminated in the market. It means that after a delay $T$ the value of mispricing is (in case the fund manager doesn't influence the market):

$$M - \mu T . \qquad (2.7)$$

However, if fund manager started to buy shares, the possible value of profit would decrease:

$$B = V(M - \mu T - \Delta p) . \qquad (2.8)$$

In this equation term $\Delta p$ corresponds to the action of the manager, and $\mu T$ – to the action of the rest of the market. It means that it is possible to neglect interactions terms between actions of the fund manager and the actions of the rest of the market.

Now, it is possible to formulate manager problem (Gabaix et al. 2003). The real goal of fund manager is a maximization of profit ($B$). If $T$ is substituted by the value of $T$ from equation (2.4) the value of $B$ become:

$$B = V(M - \mu \frac{aV}{\Delta p} - \Delta p) , \qquad (2.9)$$

where $a$ is a coefficient proportionality in the equation (2.5). Maximization of $B$ from formula (2.9) leads to the following relation:

$$\Delta p \sim V^{\frac{1}{2}} . \qquad (2.10)$$

From (2.10) it follows that

$$\xi_r = 2\xi_V \tag{2.11}$$

where $\xi_r$ and $\xi_V$ are critical indexes for distribution of returns and volumes.

The time to execution could be calculated as:

$$T \sim \frac{V}{\Delta p} \sim V^{\frac{1}{2}} \tag{2.12}$$

and the number of trades is proportional to the time:

$$N \sim T \sim V^{\frac{1}{2}}. \tag{2.13}$$

Exactly like for equations (2.13), it is immediately follows that

$$\xi_N = 2\xi_V. \tag{2.14}$$

## 2.4. Derivation of the value of critical index for the distribution of volume

Derivation of the value $\xi_V$ is based on the following 3 conditions (some of them are derived or discussed earlier):

Zipf's law is correct for mutual fund sizes ($P(s > x) \sim x^{-\xi_s}$), where s – the value of managed assets, $P$ – probability distribution of managed assets and $\xi_S \approx 1$.

The price impact (dependence of return from volume) follows formula (2.10), derived earlier.

In average, mutual funds trade in volumes proportional to the size of assets under mutual fund management:

$$V \sim S^\delta, \tag{2.15}$$

where $\delta > 0$.

The distribution of trading volumes should be such that on average mutual funds with different size should not have advantage over each other (each should have similar return, or pay similar transaction costs).

In order to measure transactions costs, the following quantity is used (Gabaix et al. 2003):

$$c(S) = \frac{\text{Annual amount lost by the fund in price impact}}{\text{Value S of the assets under management}}. \qquad (2.16)$$

According to the idea described above $c(S)$ should be independent on $S$ for big values of $S$. In (Gabaix et al. 2003) the independence of $c(S)$ from $S$ interpreted as evolutionary "survival constraint".

Loss for a block trade $V$ is equal to $V\Delta p$. Accoriding to (2.10), $V\Delta p$ is proportional to $V^{\frac{3}{2}}$. Let's call F(S) fund's annual frequency of trading. The relative annual transactions cost (in relation to the value of assets under the fund management) is equal to:

$$c(S) = \frac{F(S)V(S)^{\frac{3}{2}}}{S}. \qquad (2.17)$$

From (2.17) is follows that $F(S) \sim \frac{Sc(S)}{V^{\frac{3}{2}}}$. Probability of block trade with the size larger than x would be determined by formula:

$$P(V > x) \sim \int_{S^\delta > x} F(S)\rho(S)dS. \qquad (2.18)$$

Where $\rho(S)$ is a probability density function for mutual funds of size $S$, and $V(S) \sim S^\delta$. In agreement with Zipf's law:

$$\rho(S) \sim S^{-2}. \qquad (2.19)$$

Combining last 4 formulas it is possible to get:

$$P(V > x) \sim \int_{S > x^{\frac{1}{\delta}}} S^{1-\frac{3\delta}{2}} S^{-2} dS \sim x^{-\frac{3}{2}}. \qquad (2.20)$$

This formula leads to $\xi_V = \dfrac{3}{2}$.

That is how thus are derived.

## 2.5. Description of Johansen-Ledoit-Sornette (JLS) model

The next model considered here is called JLS model (Sornette and Johansen 2001) and it describes dynamics of bubbles. In order to create model, the description of market was simplified and considered to consist of two groups of participants (value investors and noise traders). Interaction of these groups leads to the new log-periodic oscillations in price. Authors of the model state that it could be used and was used to predict crashes and bubbles in real time.

Now, we will follow the derivation of their formula:

$$\ln E[p(t)] = A + B(t_c - t)^m + C(t_c - t)^m \cos(\omega \ln(t_c - t) - \phi), \qquad (2.21)$$

where E[…] describes expected value, $p$ - price, $t$ - time, $t_c$ – time of crash. Derivation starts from the dynamics of the price equation:

$$\frac{dp}{p} = \mu dt + \sigma dW - k dj, \qquad (2.22)$$

where $W$ – is the Wiener process, $j$ - connected with possibility of crash, $j = 0$ – for absence of the crush, and $j = 1$ after the crash, term $dj$ corresponds to the discontinuous jump where crash occurs. Jumps described by the crash hazard rate $h(t)$, where $h(t)dt$ is the probability that the crash occurs during the time $\Delta t$. Because of it:

$$E_t[dj] = h(t)dt. \qquad (2.23)$$

According to the described model herding behaviour of noise traders leads to:

$$h(t) = B'(t_c - t)^{m-1} + C'(t_c - t)^{m-1} \cos(\omega \ln(t_c - t) - \phi'). \qquad (2.24)$$

The non-arbitrage conditions could be described as

$$E_t[dp] = 0. \qquad (2.25)$$

From here it follows that $\mu(t) = kh(t)$. Eventually, it leads to the formula (2.21).

## *2.6. Description of volatility models*

The modern theory of finance is based mostly on the portfolio theory (Sharpe 1985), (Elton and Gruber 1981). However, there are many facts (Engle 1982), (Cont 2001) and theories (Engle 1982), (Cont 2001), (O'Hara 1995), (An Introduction to Technical Analysis 1999) that do not follow from this framework. The GARCH model (Engle 1982) is such model. It is supported by evidence and accepted by finance professionals and in finance literature. However, it is unclear how and from what principle GARCH could be derived. In order to describe this model let's introduce different variables that describe market movement. The most common equation used for modeling in finance is the generalized Wiener process:

$$dP = \mu_0(t)Pdt + \sqrt{h_0(t)}Pdz, \tag{2.26}$$

where $dP$ is the change in value of a security $P(t)$, $t$ is time, $dz$ is the Brownian motion, $\mu_0(t)$ is the rate of return, and $h_0(t)$ is the variance. The second set of variables considered here is similar to (Engle and Patton 2001). For this approach, $\mu_1(t)$ and $h_1(t)$ will be determined over discrete intervals $\Delta t$ in the following way: let us denote as the continuously compounded return of $P(t)$ for the time period $\Delta t$:

$$\mu_1(t) = lnP(t) - lnP(t - \Delta t) \tag{2.27}$$

We define as in (Engle and Patton 2001) conditional return and conditional variance:

$$m_1(t) = E_{t-\Delta t}[\mu_1(t)] \tag{2.28}$$

and

$$h_1(t)E_{t-\Delta t}[((\mu_1(t) - m_1(t))^2], \tag{2.29}$$

where $E_{t-\Delta t}[x(t)]$ is the expectation of variable $x$ at time $t$ with information at time $t - \Delta t$.

With unconditional variables defined as $\mu$ and $h = \sigma^2$ the GARCH (p, q) model states that:

$$h_1(t) = \omega + \sum_{i=1}^{p}\alpha_i(\mu_1(t-i*\Delta t)-\mu)^2 + \sum_{j=1}^{q}\beta_j h_1(t-j*\Delta t), \qquad (2.30)$$

where $\omega$, $\alpha_i$, and $\beta_j$ are constants.

Garch model is one of the models that implements stylized fact volatility clustering, discussed above.

# 3. Suggested framework

There are many different paradigms in finance and economics. For example, there are different paradigms of economics (classical, Marshall, Keynes, Walras, Austrian, Marxism), paradigm of finance and investment, paradigm of market microstructure. It is possible to suggest a paradigm that could combine all 3 discussed above models of econophysics. This paradigm should combine elements of non-arbitrage models for description of market impact functionality and describe reaction of market to the behaviour of big firms, when they are moving the market. Also this framework should provide explanation for behavior of market near critical points.

This framework is based on the following assumptions:

For majority of situation market follows non-arbitrage conditions.

For small number of situation market influenced by big firms that move markets.

This combination could be described using probability analysis, where small probability would correspond to big market moves.

Because observation of big movers could lead to arbitrage opportunities, this framework could be a theoretical foundation for possibility of the technical analysis.

If the action of the big firm could lead to the arbitrage opportunity, small-firm could self-organize in order to take advantage of this opportunity. The time for this self-organization is substantially bigger than the time of action for one big firm.

This framework could be applied for description of market impact, probability laws of big market move, crashes and the description of volatility.

# 4. Application to the theory of market impact

It is possible to apply ideas of proposed framework to the description of market impact. In order to do that, the equation (2.1) needs to be modified in such way that it would include 2 additional terms: the first term, which describe the possibility that market move is connected with big financial firm and another term describing the possibility of market move connected with some deterministic changes (a la technical analysis), Specifically, equation (2.1) would look like:

$$(1 - p_A - p_B)(p\Delta S_1 + (1-p)\Delta S_2) + (p_A + p_B)(\Delta S - \Delta S_3) = 0, \qquad (4.1)$$

where

$p_A$ is a probability that market move was determined by big financial players

$p_B$ is a probability of a market move that is deterministic and determined by the change in the composition of big players (it is deterministic but unexpected by the majority of market)

$p$– probability that market expected change $\Delta S_1$

$1 - p$ – probability of unexpected market change $\Delta S_2$

$\Delta S$ – the value of future price change

$\Delta S_3$ – the value of future price change if it is determined by the big players.

Equation (4.1) contains 2 different terms:

The first term is exactly the same as in the model of Bouchaud and al. and corresponds to the behaviour of normal market.

Second term $(p_A + p_B)(\Delta S - \Delta S_3)$ corresponds to situation when market is determined by the behaviour of big players. In that case the possible future change $\Delta S$ is equal to the change originated from the behaviour of the big market firms $\Delta S_3$.

Let's repeat that $\Delta S_1$, $\Delta S_2$ and $\Delta S_3$ are different price changes in the market:

$\Delta S_1$ – expected market change in the normal market,

$\Delta S_2$ – unexpected market change in the normal market,

$\Delta S_3$ – market change when big players move market.

The dominance of terms 1 and 2 could be connected with level of liquidity. When level of liquidity in the market is high enough, market influence of big players is absorbed and their influence could be described only in probabilistic way (as the first term in (4.1). However, when the level of liquidity is low, any action of big players would propagate without resistance (and the second term in (4.1) became dominant).

# 5. Application of the new framework to the behaviour of big players in the market

In description of a new framework new features were added to the description of market movements caused by big players (Gabaix et al. 2003). Because of it, it is possible to modify distribution obtained by Stanley and al., adding smaller terms. So, which new players or contributions added?

1. Big deterministic changes connected with both the unexpected change of population and composition of big players (I1)

2. Small deterministic changes connected with expected change of population of big and small players (I2)

3. Changes in unexpected random behaviour of market (I3)

It is assumed here that

$$I1 \gg I2 \gg I3 \qquad (5.1)$$

It is possible to suggest that value of $I1$ in the first approximation could be described by the least action principle:

$$\int_{t_1}^{t_2} L(S, \dot{S}, t) dt -> \min . \qquad (5.2)$$

Basically, because most of changes are predictable, and some unpredictable changes are small changes by amplitude with time, the formula (5.2) means that additional important changes (to addition to distribution of changes, proposed by Stanley and al.) connected with minimal possible profit. This additional minimal possible profit appears because traders that observed the market moved market in such way that possible profit becomes minimal.

The formula (5.2) states that minimal possible profit $L$ in unit of time depends on $S$ (price of instrument), its derivative over time and time. It is assumed here that exist non-random component in price changes and it is possible to define derivatives.

Because this formula is exactly the formula that describes movements in classical mechanics, it is possible to suggest that as in classical mechanics dependence of $S$ over time could be described by formula:

$$S = S_0 + V_0 t + \frac{at^2}{2} \quad (5.3)$$

(it is assumed here that the first and second derivatives of $S$ over $t$ are constants), or it could be rewritten as:

$$\Delta S = V_0 \Delta t + \frac{a \Delta t^2}{2} \quad (5.4)$$

Let's derive distribution for $P_1(\Delta S)$. Under $P_1(\Delta S)$ we understand the distribution of $\Delta S$ caused by unexpected changes in population and composition of big players. Because $P_1(\Delta S)$ is the same value for all times, it is possible to deduce that for $V_0 \Delta t \gg \frac{a \Delta t^2}{2}$ $P_1(\Delta S)$ has the same value, or

$$P_1(\Delta S) \sim \Delta S^{-\xi_{r1}}, \quad (5.5)$$

where $\xi_{r1} = 0$. If $V_0 \Delta t \ll \frac{a \Delta t^2}{2}$

$$P_1(\Delta S) \sim \Delta S^{-\xi_{r2}}, \quad (5.6)$$

where $\xi_{r2} = 0.5$.. It means that for small $\Delta S$ $\xi_{r1} = 0$, for bigger $\Delta S$ $\xi_{r2} = 0.5$, and for large $\Delta S$ values $\xi_r = 3$, according to Stanley and al. (Gabaix et al. 2003) analysis.

Now we could suggest analysis that could describe combined distribution for all values $\Delta S$. It is based on the assumption that combined distribution could be explained on the basis on the second order phase transition. Specifically, in order to use the second order phase transition theory it is necessary to have an order parameter. Here, it is assumed that an order parameter is the inverse ratio of probability of stock market changes caused by behaviour of big players over probability of changes caused by changes in composition of big players. The theory of phase transitions of the second order is described earlier in this book. The order parameter $\frac{1}{P}$ is described by the scaling function

$$\frac{1}{P} = A\Delta S^{\xi_r} f(h\Delta S^{\xi-\xi_r}). \tag{5.7}$$

Scaling function $f(z)$ for 2 specific cases behaves in the following way:

$$\begin{aligned} f(z) &\to \{z^\beta, z \gg 1\} \\ f(z) &\to \{const, z \ll 1\} \end{aligned} \tag{5.8}$$

# 6. Application of the new framework to the theory of crashes

As for the other areas of finance, the idea of suggested framework could be applied for the description of market crashes. The idea of that application based on the interaction between 2 groups: big players and small participants (noise players) in the market during period before market crash. According to the approach, pioneered by Sornette and al. (Sornette and Johansen 2001) one of the parameters, important for prediction of the market before the crash is $h(t)$. We suggest that in order to consider interaction of 2 group players, we need to consider not one value of $h(t)$, but 2 values: $h_1(t)$ and $h_2(t)$, where $h_1(t)$ describes probability of crash by small players (regular market), and $h_2(t)$ parameter that determine probability of crash for big players (firms, which could move the market). The equation (3.2.24) could be rewritten as:

$$E_t[dj] = (C_1 h_1(t) + C_2 h_2(t))dt . \tag{6.1}$$

It is assumed here that both $h_1(t)$ and $h_2(t)$ are not independent, but connected with each other. The ratio $\frac{h_1}{h_2}$ should depend on the perceived distance to the crash. The principle that could be used for determining ratio $h_r = \frac{h_1}{h_2}$ could be the least action principle for minimizing possible profit for observers of crash:

$$\int L(h_r, \dot{h}_r, \Delta) d\Delta -> \min , \tag{6.2}$$

where Δ is the distance to the crash.

The justification for this formulation could be based on the interaction between small and big players, small players react to the actions of the big players and opposite: big players react to the actions of small players. These interactions over time lead to quasi-deterministic behaviour. The interaction happened in such way that participants happened to earn a minimal possible profit described by the equation (6.2). Assuming the ratio $h_r$ is oscillating around stable value (similar to the behaviour of oscillator in the classical mechanics), it is possible from (6.2) to obtain equation:

$$\frac{d^2h}{d\Delta^2} + \omega^2(h - h_c) = 0. \tag{6.3}$$

The solution to this equation would be:

$$h = h_c + h_{r0}\cos(w\Delta). \tag{6.4}$$

In order to find solution for $h_r$, it is necessary to have assumption about $\Delta$. Let's consider 2 different classes:

1. $\Delta$ is decreasing as $t$ is approaching $t_c$:

$$\Delta \sim (t - t_c)^\alpha, \alpha > 0, \tag{6.5}$$

2. $\Delta$ is increasing as $t$ is approaching $t_c$:

$$\Delta \sim \ln(t - t_c) \text{ or } \Delta \sim (t - t_c)^{-\alpha}, \alpha > 0. \tag{6.6}$$

Solution h for $\Delta \sim \ln(t - t_c)$ would lead to the solution, suggested by Sornette and al. Solutions with $\Delta \sim (t - t_c)^\alpha, \alpha > 0$ and $\Delta \sim (t - t_c)^{-\alpha}, \alpha > 0$ would lead to the new solutions near crash.

# 7. Application of the new framework to the description of volatility

Let's describe additional assumptions connected with framework suggested by Bouchaud and al. (Bouchaud et al. 2009). Specifically, to the assumptions described in this framework, let's add the following one:

Expected change of stock price made by knowledgeable people, and unexpected change caused by noise traders. Under this assumption the $p$ in equation (2.1) :

$$p\Delta S_1 + (1-p)\Delta S_2 = 0 \tag{7.1}$$

corresponding to probability of movement caused by knowledgeable traders, and $1-p$ to the probability that movement caused by noise traders. According to this assumption it is possible to formulate the following theorem:

The dominant contribution to the volatility caused by the noise traders.

This theorem has a very simple proof. According to the suggested assumption the part of the volatility term caused by noise traders:

$$\sigma_2^2 = (1-p)\Delta S_2^{\,2}, \tag{7.2}$$

where the same notations used as in the formula (7.1). Than the contribution to volatility by knowledgeable traders is:

$$\sigma_1^2 = p\Delta S_1^{\,2} \tag{7.3}$$

and the summary volatility $\sigma$ could be calculated as:

$$\sigma^2 = \sigma_1^2 + \sigma_2^2. \tag{7.4}$$

Let's compare $\sigma_1$ and $\sigma_2$. Because of (7.1) $\Delta S_2 = \dfrac{-p}{1-p}\Delta S_1$, and it follows that

$$\sigma_2^2 = \dfrac{p}{1-p} p\Delta S_1^{\,2}. \tag{7.5}$$

Because $\frac{p}{1-p} \gg 1$ it follows that $\sigma_2 \gg \sigma_1$.

It means that the less informative members of the trader population are going to determine the value of volatility. This theorem could explain the difference between values of actual and implied volatility in the theory of option pricing. It also could lead to doubts about foundations for existing theories of volatility. The similar analysis could be performed in the framework of big movements (Gabaix et al. 2003). And that analysis could lead to the similar results. The terms corresponding to the bigger changes in the market prices lead to the bigger contributions to volatility. Because the largest contributions to the volatility connected with non-informative population and processes, it is possible to assume that their behaviour could be described by dynamic processes.

In this chapter we will show that the GARCH model could be derived from the least common action principle. The least common action principle was successfully applied in different areas of science, including economics (Magill 1970), control theory (Sage and White III 1977), and physics (Landau and Lifshitz 1976 – 1981). Some economic reasoning that leads to the least action principle is also provided in this chapter. The set of possible ideas that could lead to volatility equations is presented in the next section (Economic Reasoning). The suggested approach could lead to a new method of volatility prediction. It allows one to obtain not only a functional dependence similar to GARCH, but also the value of coefficients, close to what is observed in practice.

In order to use the advantages of continuous functions, let's introduce new functions $h_2(t)$ of volatility over time constructed in a specific way (and call them quasicontinuous). These functions $h_2(t)$ are continuous in time, such that they coincide with $h_1(t)$ for time sequences $0, \pm \Delta t, \pm 2\Delta t, \ldots$ For example, it is possible to use splines to construct such functions with the first and second derivative (Press et al. 1992), and to show, with additional conditions, that these functions exist and are unique.

All equations and predictions are specified using the $h_2(t)$ functions. We consider $h_2(t)$ to be a "good enough" approximation for the real function $h_0(t)$, specifically for time intervals larger than $\Delta t$.

We will suggest an equation describing the behavior of $h_2(t)$, find its solution, and then, considering that $h_2(t)$ is an approximation for $h_0(t)$, use it to write formulas for $h_0(t)$. The difference between $h_0(t)$, $h_1(t)$ and $h_2(t)$ is presented in Table 1.

Table 1. Different variances considered in this article.

| Dispersion | Description |
|---|---|
| $h_0(t)$ | Continuous |
| $h_1(t)$ | Discrete over the period $t$-$\Delta t$ to $t$ |
| $h_2(t)$ | Continuous coincides with $h_1(t)$ at times ..., $t$-$\Delta t$, $t$, $t$ +$\Delta t$,... |

### *7.1. Economic reasoning*

Here, two different versions of economic reasoning are considered. The first one is a risk-neutral approach.

There is a small additional premium $dP_V$ connected with changing volatility. In continuous coordinates:

$$dP = \mu_0(t)Pdt + dP_V + \sqrt{h_0(t)}Pdz, \qquad (7.6)$$

where $dP_V$ is proportional $dt$ and could be described as:

$$dP_V = L_A dt, \qquad (7.7)$$

if it's assumed that $h_2(t)$ functions have a "good enough" modeling description of the world, it is possible to substitute $h_0(t)$ in equation (7.6) with $h_2(t)$ and suggest that $L_A$ is only a function of $h_2(t)$, $\overset{*}{h_2}(t)$, and $t$:

$$dP = \mu_0(t)Pdt + L_A(h_2, \overset{*}{h_2}, t)dt + \sqrt{h_2(t)}Pdz. \qquad (7.8)$$

If there is a competition between different dealers for the option contracts, we could also suggest that the premium $dP_V(t)$ over time tends to be minimum or:

$$\int L_A(h_2, \overset{*}{h_2}, t)dt -> \min \qquad (7.9)$$

The fact that there is a risk-neutral measure means that there is a way to calculate the price for an option contract; however, it doesn't influence the volatility prediction.

For this consideration, let's assume that arbitrage exists. It is very small, short-lived, and dynamic. These ideas are similar to the ones considered in the work of Adamchuk and Esipov (Adamchuk and Esipov 1997). For this case we assume that equation (2.25) holds. In addition, it is assumed that even for the best hedged portfolio $P_I$ arbitrage exists and could be measured using the following quantity:

$$\int (dP_I - rP_I dt)^2, \qquad (7.10)$$

and we denote $L$ as $L_B$, where $L_B dt = (dP_I - rP_I dt)^2$. If we assume, that arbitrage tends to minimum, and $L_B$ depends on $h_2(t)$, $\overset{*}{h_2}(t)$, and $t$, we could write it as in the first consideration:

$$\int L_B(h_2, \overset{*}{h_2}, t) dt \to \min. \qquad (7.11)$$

## 7.2 Equations for volatility

Now we will derive an equation for $h_2(t)$ which would be identical for both considerations, and we substitute $L_A$ and $L_B$ with $L$ and assume that $L$ has the following properties:

$$\int L(h_2, \overset{*}{h_2}, t) dt \to \min, \qquad (7.12)$$

From this equation (7.12) we could obtain the equation of Euler-Lagrange (Landau and Lifshitz 1976 – 1981):

$$\frac{d}{dt} \frac{\partial L}{\partial \overset{*}{h_2}} = \frac{\partial L}{\partial h_2}. \qquad (7.13)$$

In order to obtain a new result we need to choose a form of dependence of $L$ from $h_2(t)$ and $\overset{*}{h_2}(t)$. For a small $h_2(t)$ and $\overset{*}{h_2}(t)$ we could have a Taylor's series for $L$. Also, because $L$ is optimal, we could assume that the first non-existent power in this series is square for $h_2(t)$ and that $L$ doesn't depend on the sign of $\overset{*}{h_2}(t)$. On the basis of these assumptions we could write that $L$ is equal to:

$$L = \frac{a \overset{*}{h_2}{}^2}{2} + \frac{b h_2{}^2}{2}, \qquad (7.14)$$

If we assume that $L$ doesn't directly depend on $t$, we could write that:

$$dL = \frac{\partial L}{\partial h_2} dh_2 + \frac{\partial L}{\partial \overset{*}{h_2}} d\overset{*}{h_2}. \qquad (7.15)$$

Using different transformations for (7.13) and (7.15), we could show that:

$$\frac{d}{dt}(\overset{*}{h_2} \frac{\partial L}{\partial \overset{*}{h_2}} - L) = 0. \qquad (7.16)$$

For $L$ described by formula (7.14), the last equation (7.16) will lead to:

$$\frac{a \overset{*}{h_2}{}^2}{2} - \frac{b h_2{}^2}{2} = c_1, \qquad (7.17)$$

where $c_1 =$ const.

If we consider $c_1 \ll \frac{b h_2{}^2}{2}$, we could obtain from (7.17):

$$\overset{*}{h_2} = \sqrt{\frac{b}{a}} h_2 (1 + \frac{c_1}{b h_2{}^2}). \qquad (7.18)$$

We need to suggest a connection between $h_2(t)$ and $h_0(t)$. Because of the way we constructed the new function $h_2(t)$, we could assume that

$h_2(t) = h_0(t)$ for $t = 0, \pm \Delta t, \pm 2\Delta t, \ldots$

We could write that:

$$\overset{*}{h_2} = \frac{h_2(t + \Delta t) - h_2(t)}{\Delta t} = \frac{h_0(t + \Delta t) - h_0(t)}{\Delta t}, \qquad (7.19)$$

$h_2(t)$ describes average volatility squared between $t$ and $t+\Delta t$. We could substitute it, using equation (7.6), with

$$h_2 P^2 \Delta t = (\Delta P - \mu P \Delta t - dP_V)^2 \qquad (7.20)$$

for the first economic derivation and equation (2.25)

$$h_2 P^2 \Delta t = (\Delta P - \mu P \Delta t)^2 \qquad (7.21)$$

for the second economic derivation.

Neglecting the second term in (7.18), using previous equation (7.19), and neglecting $dP_V$ in the right part of (7.20), we could obtain:

$$\frac{h_0(t+\Delta t) - h_0(t)}{\Delta t} = \sqrt{\frac{b}{a}} h_2, \qquad (7.22)$$

and

$$h_0(t+\Delta t) = h_0(t) + \sqrt{\frac{b}{a}} \Delta t \left(\frac{\Delta P}{P\Delta t} - \mu\right)^2, \qquad (7.23)$$

which is very similar to the GARCH equation.

### 7.3. Summarized assumptions

Let's summarize the assumptions used in (7):

There is a correspondence between real volatility and a new (continuous) volatility $\sqrt{h}$ (and $\sqrt{h}$ is a good approximation for the prediction of a real volatility).

It's possible to introduce function $L$, dependent on a new volatility, such that $\int L dt -> \min$.

This function $L$ depends on the continuous volatility $\sqrt{h}$, its derivative over time $t$, and $t$.

Because of its optimality and symmetry, $L$ is equal to $\dfrac{a h^{*2}}{2} + \dfrac{b h^2}{2}$.

These assumptions are presented in Fig.1:

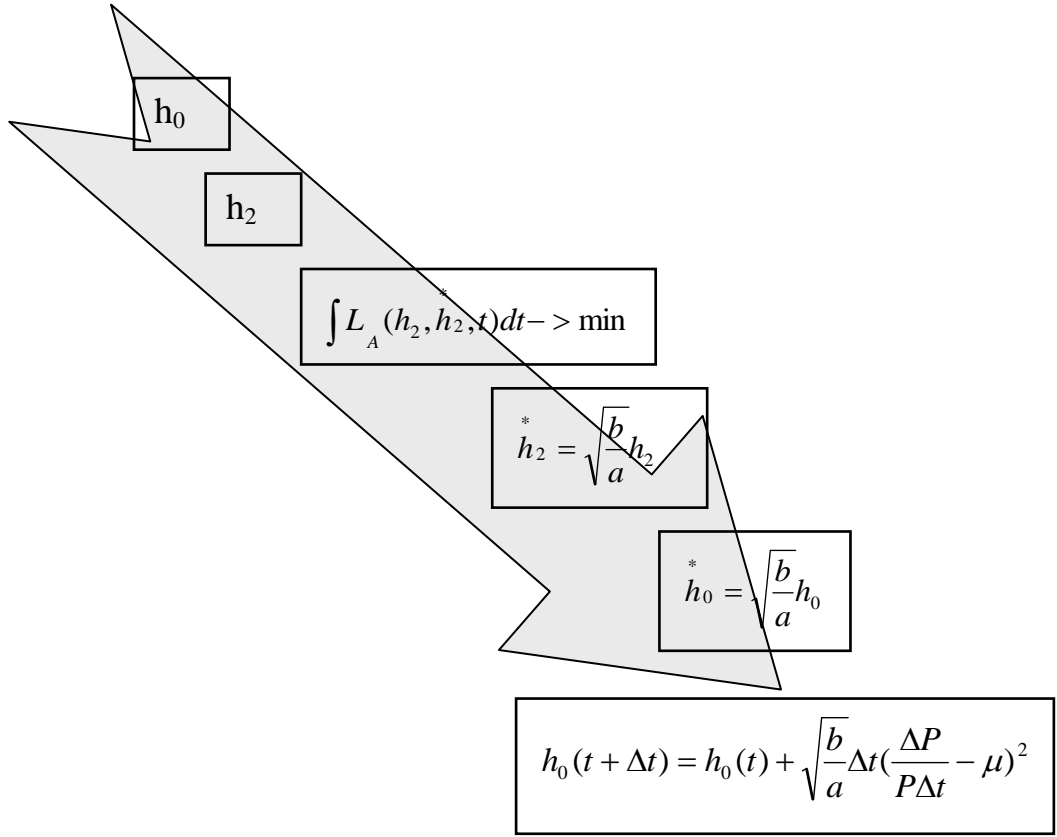

Fig.1. Graphical description of assumptions and results.

## 7.4. Comparison with data and possible future development

Now let's compare the results of this modeling with the coefficients of GARCH(p, q) determined in (Engle and Patton 2001). First of all, it appears, the best correspondence with analyzed data exists for GARCH(1, 1), which has the closest resemblance to formula (7.23). It appears that the coefficients for equation (7.5), determined from real data, are very close to the ones suggested by formula (7.23):

$\omega$ is close to 0, $\alpha$ is close to 1 (see table 2 from (Engle and Patton 2001)). This means, that the suggested approach could predict not only the form of dependence, but the coefficients too. This could be used as the basis of a new methodology of volatility prediction.

Table 2. Comparison between GARCH(1,1) coefficients from (Engle and Patton 2001) and present results.

| From article (Engle and Patton 2001) | Current chapter |
|---|---|
| 0.0082 | 0 |
| 0.9505 | 1 |

# 8. Conclusion

We call proposed models quasi-microstructure models. They are similar in some respect to the standard market microstructure models, but in difference to usual microstructure models they are concerned not with structure of the market players (noise traders, strategic traders, market makers), but with the strength of market players (players, that move markets, and players, that are not).

# References.